\documentclass[letterpaper]{article} 
\usepackage{aaai25}  
\usepackage{times}  
\usepackage{helvet}  
\usepackage{courier}  
\usepackage[hyphens]{url}  
\usepackage{graphicx} 
\usepackage{natbib}  
\usepackage{caption} 
\usepackage{algorithm}
\usepackage{algorithmic}

\usepackage{xcolor}
\newcommand{\answerYes}[1]{\textcolor{blue}{#1}} 
 
\newcommand{\answerNA}[1]{\textcolor{gray}{#1}} 
 
\usepackage{newfloat}
\usepackage{listings}
\DeclareCaptionStyle{ruled}{labelfont=normalfont,labelsep=colon,strut=off} 
\floatstyle{ruled}
\newfloat{listing}{tb}{lst}{}
\floatname{listing}{Listing}
\title{TikTok Search Recommendations: Governance and Research Challenges}
\author{
    Taylor Annabell\textsuperscript{\rm 1},
    Robert Gorwa\textsuperscript{\rm 2},
    Rebecca Scharlach\textsuperscript{\rm 3},
    Jacob van de Kerkhof\textsuperscript{\rm 1},
    Thales Bertaglia\textsuperscript{\rm 1}
}
\affiliations{
    \textsuperscript{\rm 1}Utrecht University, Utrecht, The Netherlands\\
    \textsuperscript{\rm 2} WZB Berlin Social Science Center, Berlin, Germany\\
    \textsuperscript{\rm 3} University of Bremen, Bremen, Germany\\
    t.annabell@uu.nl, robert.gorwa@wzb.eu, scharlach@uni-bremen.de, j.j.w.vandekerkhof@uu.nl, t.f.costabertaglia@uu.nl

}

\begin{document}

\maketitle

\begin{abstract}
Like other social media, TikTok is embracing its use as a search engine, developing search products to steer users to produce searchable content and engage in content discovery.  Their recently developed product \textit{search recommendations} are preformulated search queries recommended to users on videos. However, TikTok provides limited transparency about how search recommendations are generated and moderated, despite requirements under regulatory frameworks like the European Union's Digital Services Act. By suggesting that the platform simply aggregates comments and common searches linked to videos, it sidesteps responsibility and issues that arise from contextually problematic recommendations, reigniting long-standing concerns about platform liability and moderation. This position paper addresses the novelty of search recommendations on TikTok by highlighting the challenges that this feature poses for platform governance and offering a computational research agenda, drawing on preliminary qualitative analysis. It sets out the need for transparency in platform documentation, data access and research to study search recommendations.
\end{abstract}

\section{Introduction}
\begin{figure}
    \centering
    \includegraphics[scale=.35]{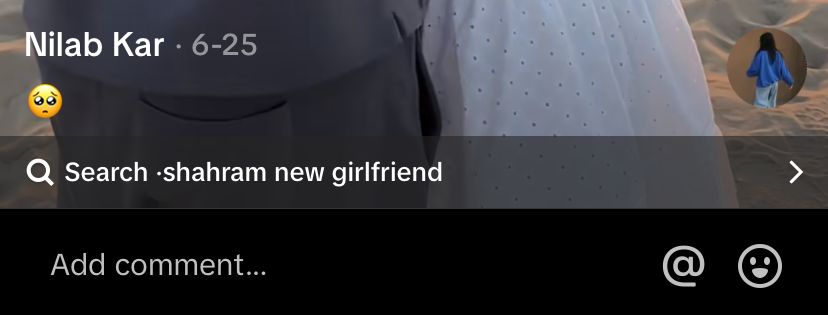}
    \caption{Search recommendation appearing on 18 July 2024 on Nilab Kar’s  TikTok video originally shared 25 June 2024.}
    \label{fig:example}
\end{figure}

Social media have long featured search and other user-driven content discovery features. But popular feed-based platforms are rolling out novel ‘search’ tools to help users find content that might be relevant to them. TikTok prompts users to engage in their search experience through search recommendations, which have also recently appeared on YouTube Shorts. Moreover, Instagram has expressed intent to develop an equivalent product soon in response to the development by TikTok. On TikTok, search recommendations are preformulated search terms that appear as an overlay on some videos (see Figure~\ref{fig:example}) or at the top of the comments section, which offer a new access point for generating search queries. Search becomes a recommended action to be taken after watching the video, and what should be searched is conceived of as a new form of recommendation.

TikTok frames search terms as a reflection of conversations surrounding the specific video. The content of comments and common searches made after watching a video are the only factors TikTok attributes to shaping recommendations, clearly positioning the platform as a mere aggregator of search terms rather than a generator.

TikTok’s approach to search through recommendations has not been matched by adequate transparency measures. There is limited platform documentation explaining how search recommendations function and a lack of clarity regarding the moderation and reporting processes. This is troubling given that TikTok’s search recommendations pose numerous governance problems, including:  potentially harmful search terms that spread misinformation, amplify drama, fuel tabloid-like speculation, and perpetuate stereotypes~\cite{lorenz_tiktok_2024}; the possibility for the opaque search algorithm to be gamed, for example to elevate political candidates, as reported during the annulled Romanian presidential election~\cite{olari_rise_2024}; a lack of recourse for users, including the video creator, to influence the visibility and content of these recommendations, which may not align with the content of the video itself. 

This position paper addresses the novelty of search recommendations on TikTok by situating it at the intersection of recommendation systems for search engines and social media platforms, as well as ongoing debates about platform companies’ responsibility regarding moderation. Drawing on preliminary qualitative analysis of search recommendations on Dutch influencers' videos, we set out (i) challenges of search recommendations for platform governance and (ii) a computational research agenda to examine this search product, reflecting on challenges for the study of search recommendations within the limits of publicly available data.  

\section{TikTok’s Orientation Towards Search}
The orientation of TikTok towards what it refers to as ``search experience'' advances and capitalises on the platform's increasing usage as a search engine. According to an Adobe Express survey~\cite{adobe_express_using_2024}, 40\% of Americans use TikTok as a search engine, echoing TikTok internal data~\cite{tiktok_introducing_2024}. The value of `social search' -- searching via social media~\cite{evans_towards_2009} -- is evident in the development of products aimed at different TikTok end-users, solidifying entanglements between search, recommendation, and content.

Creators are now encouraged to develop search-friendly content due to the increased use of search on TikTok. Creator Search Insights show creators the topics that are being searched for, the ``content gap topics'' (popular searched topics that have few videos associated with them) and search metrics for their content, such as how often posts were viewed on search results and what percentage of views come from the search feed. Furthermore, the Creator Rewards Program, which allows eligible creators to collect rewards on qualifying content, considers `search value' as one of the four key factors in its rewards formula. This explicitly incentivises content production that aligns with high-demand search topics or provokes searches, linking searchability with monetisation and indicating a paradigm shift in content production from optimisation for FYP recommendation to \emph{search} recommendation. In doing so, TikTok attempts to manage the supply-demand issue with search: delivering `relevant' videos to meet search queries.

When it comes to TikTok's advertising business model, ads can be displayed on the TikTok search results page, illustrating how search fits in the platform's multisided marketplace. Automatic Search Placement allows existing ad content to appear against relevant user queries, while Search Ads Campaign provides more control through keyword-based ads, enabling advertisers to target specific search results pages. The Keyword Suggestion Tool recommends keywords for these search ads to assist optimisation, while also offering monthly impression predictions. Each keyword submitted in these ad campaigns undergoes moderation against ``legal requirements and TikTok policy standards''~\cite{tiktok_keyword_2025} with industry-specific guidelines. Consequently, the documentation for advertisers indicates some level of oversight regarding the use of search via keywords for targeting advertising. In this advertising pipeline, TikTok sets boundaries of acceptability and offers transparency on keyword rejections, allowing advertisers to appeal platform decisions.

Users access search via the search bar on the homepage, by tapping hashtags in content captions, or by engaging with search recommendations. These recommendations appear in multiple locations: at the top of a video's comments section, below the video, highlighted within comments, and in the video detail page's search bar. Regardless of whether a user creates their own search query or utilises a pre-existing search recommendation that is served to them via FYP, results are algorithmically generated based on factors such as relevance and engagement with search interventions provided for specific topics.

Search recommendations offer a new way for users to interact with TikTok by presenting preformulated queries. Similar to hashtags, search recommendations can be viewed as ``searchable talk''~\cite{zappavigna_searchable_2015}, but they provide an access point to search that isn't enabled by social tagging. We lack empirical insights into how search recommendations are generated and what queries are recommended on TikTok videos. Motivated by this research gap, we briefly introduce a small-scale qualitative study, which informs the computational social science agenda we propose in section 3.

\begin{figure}
    \centering
    \includegraphics[scale=.8]{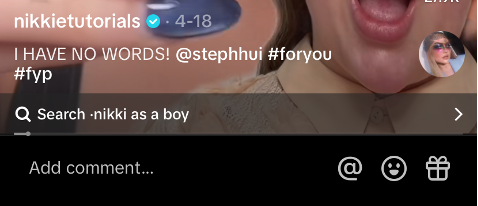}
    \caption{Search recommendation appearing on 17 July 2024 on NikkieTutorials's TikTok video originally shared 18 April 2024.}
    \label{fig:example2}
\end{figure}

Based on a curated dataset~\cite{gui2024across}, we identified the 10 most-followed Dutch influencers on TikTok and collected search recommendations from videos on their accounts between January and July 2024, tracking changes over three consecutive days. We observed that the presence of search recommendations and their content was dynamic, resulting in the collection of 92 search recommendations under videos and 167 search recommendations in the comments section of 110 videos. Qualitative thematic analysis of the content reveals that personal lives and identities are made searchable and subject to speculation. This ranges from queries that can be considered less harmful -- languages spoken or full name -- to contextually problematic -- gender identity. For example, the search recommendation ``Nikki as a boy'' appears on two videos of trans influencer NikkieTutorials (see Figure~\ref{fig:example2}), which can be interpreted as transphobic. For some creators, recommendations often involve speculation about their current and previous romantic relationships, bringing up names and characters not included in the creators' content, which serves to amplify drama and push false narratives. Additionally, some recommendations included specific product names, which encouraged consumer-based queries, intensifying the blurred boundaries between entertainment and commerce on TikTok. Working with this small dataset raises questions about the values that underlie its generation and highlights the significance of context in understanding the implications of how search recommendations may steer users towards queries that amplify topics, issues, and products, within the power imbalance between the platform and the user.

Prompted queries invert the ``key technological affordance of search engines \ldots{} their open-ended nature''~\cite{graham_investigating_2023} because TikTok determines what is searchable, and users cannot modify or disable search recommendations. The narrative presented by TikTok suggests that recommendations originate from users, implying the platform positions itself as a neutral intermediary that merely aggregates conversations about a video. But the method through which they index specific aspects of a video -- its description, comments associated with it, or potentially even the content of the video itself -- is opaque, strategic, and possibly political. 

\section{Governance Challenges}
Search recommendations present several challenges for platform governance. These are related to, and yet nonetheless unique, reconfigurations of classic issues that platforms moderating content, search results, and recommender systems face. The most obvious of these are transparency problems. Firstly, TikTok provides almost no explanation about what search recommendations are and how they are generated. In the Community Guidelines~\cite{tiktok_tiktok_2024}, TikTok establishes how search queries may be restricted when they violate rules and acknowledges the presence of recommendations:

\begin{quote}
    We provide search suggestions that are relevant to you. These search suggestions can be found throughout the platform, including autocomplete in the search tool, content from the You May Like section, and while watching content in the FYF.
\end{quote}

Search suggestions encompass both search recommendations and autocomplete, with the proactive search recommendation downplayed and not described in any technical or policy-relevant detail including the support page, How TikTok recommends content~\cite{tiktok_how_nodate}. It only states that the search feature ``also recommends search terms for you to discover content''.

Some detail is provided via TikTok Creator Academy webpages, which ``provides transparency into how TikTok works'' for \emph{creators}~\cite{tiktok_tiktok_2025}: recommendations are generated using AI based on a ``variety of factors,'' two of which are user comments on a video and what other users search for after watching. In doing so, TikTok offers more access to creators than other users. This documentation suggests that search recommendations can be solely attributed to users, following Google's discursive approach of remaining vague and contributing to the ``widespread misconception'' that Google Autocomplete reflects the most searched queries~\cite{graham_investigating_2023}. While TikTok implies that search recommendations do not index video, it is unclear whether audio transcripts or descriptions might be additional factors.

Secondly, the way TikTok moderates search recommendations remains unclear. Within the TikTok Creator Academy, safety in search is acknowledged (unlike in user-facing platform documentation), which includes the application of Community Guidelines to search results and queries. This extends to search recommendations, according to a TikTok spokesperson, which aimed to filter out harmful suggestions, such as bullying phrases. News reports indicate concerns from creators about recommendations related to weight gain and loss~\cite{schroeder_is_2023}, HIV and LGBTQ+ issues~\cite{lorenz_tiktok_2024}, and gender identity~\cite{crimmins_is_2023}, suggesting potential gaps in moderation policies and enforcement regarding `harmful' recommendations. The sexist and homophobic stereotyping reflected in these examples aligns search recommendations with critiques of Google Search~\cite{noble_algorithms_2018} and raises the issue of context.

Thirdly, users have limited control over how search recommendations are served. The TikTok Creator Academy states that users can report search results and recommendations, but our analysis of reporting search recommendations in the interface shows that recommendations appearing at the top of the video's comments can be reported, while those below the video cannot. By holding down the recommended search in the comments, users can choose from a menu that includes options such as `not interested,' `unrelated content,' `inappropriate content'. Once users click on a search recommendation, the three-dot menu next to the search includes a function to ``share feedback, " which is limited to the results rather than the query. This differs from the broader range of options available under community guidelines when reporting search autocomplete suggestions.

To prevent specific words from appearing in search recommendations, creators can utilise comment filters. While this solution is publicly acknowledged by TikTok~\cite{biino_tiktoks_2024} it is missing from any official documentation. The information presented regarding this tool on TikTok Creator Academy, Safety Centre, Support webpages, Community Controls, and Community Guidelines states that creators are responsible for managing the comments received on their videos (and therefore search recommendations as well).

These transparency concerns may pose issues for TikTok under regulatory frameworks such as the Digital Services Act (Regulation 2022/2065, `DSA'). TikTok is required to provide transparency on recommendation factors under DSA Article 27(1) concerning recommender systems, which arguably includes these search recommendations, and we believe that existing platform documentation fails to achieve this. Similarly, the design of recommender interfaces must not deceive or manipulate users in a manner that impairs their decision-making ability; so-called `dark patterns' are prohibited under Article 25(1). Since search recommendations can promote products, and are extremely opaque and unpredictable, they raise concerns regarding manipulative commercial practices.

Conversely, if we view search recommendations as `information', Article 16 requires TikTok to enable users to submit notices to report information they deem `illegal content'. As indicated by the examples we present, search recommendations may contain controversial content, for which users have limited means to report as `illegal'. This raises a compliance question under Article 16(1) DSA, especially as a creator whose content might include search recommendations is not notified of the content of those recommendations.

While this is still an emerging area and industry practices are somewhat in flux, we believe that TikTok's search recommendations are an important research object that needs to be better understood and more closely scrutinised, especially in light of broader adoption of such recommendations by, for example, Instagram. However, the current affordances of search recommendations -- and the particular design of TikTok -- also make them especially difficult to study systematically. 

\section{Computational Challenges for Research}
Studying TikTok's search recommendations at scale presents several methodological and infrastructural challenges. Data access is a key limitation in conducting large-scale empirical studies. Within the existing TikTok Research API access, search recommendations are not included in the available data. Moreover, the two data sources that TikTok has identified as key for its search index, user comments and search queries, are only partially accessible.

In addition to data access constraints, recommendations are embedded within a constantly shifting interface, where their position and visibility change across users and time in a way that is generally unusual for a core platform feature. Recommendations are also deeply contextual: what appears as a neutral query may carry different meanings depending on the creator's identity, the video's content, and the broader social discourse. Computational research designs must, therefore, navigate both the opacity of the system and the sociotechnical complexity that drives and shapes recommendations. We identify four core challenges for computational research into TikTok's search recommendations. We present them not as prescriptive tasks but as starting points for a broader research agenda, each pointing to conceptual methodological challenges that must be addressed to understand the implications and potential harms of search recommendations at scale. We contextualise each challenge with an example and highlight the empirical and computational complexity it involves.

\paragraph{Understanding how recommendations are generated} TikTok frames recommendations as emerging from user conversations, yet it remains unclear which specific comment features shape which search terms are generated and when. In our dataset, videos by Dutch influencer Nilab Kar had recommendations related to her personal relationships, such as "shahbram new girlfriend" and "nilab kar and her ex" as recommended searches, even though the content itself did not mention her current or past partner. On one day of our data collection, 18 out of 23 recommended queries on her videos referenced personal relationships, suggesting a strong correlation with speculation in the comments.

Computationally, this challenge involves modelling the relationship between the comment section and the search recommendation. Relevant features might include topic, sentiment, engagement metrics (likes, replies), recency, and user identity. Yet, due to the black-box nature of the system, it is difficult to attribute causality or determine which signals are most influential. More broadly, this raises questions about how visibility is distributed and how speculation is legitimised through algorithmic curation.

\paragraph{Detecting when coordinated behaviour affects recommendations} Coordinated commenting can shape search recommendations, even across unrelated videos. During the 2024 Super Bowl halftime show, users repeatedly commented phrases like ``Bisan and Motaz would absolutely crush this'' on entertainment content, leading to ``Bisan and Motaz NFL'' appearing as a recommended search term~\cite{dibenedetto_did_2024}. This attempted to bring attention to the accounts of Palestinian citizen journalists~Bisan Owda and~Motaz Azaiza. Detecting such coordination is challenging; It requires identifying semantic and temporal patterns across large volumes of comments and across different videos. The relevant discourse may often originate outside the platform or video itself. This challenge is fundamental for content moderation: understanding whether recommendation systems can be manipulated and what signals are (or are not) monitored is key to assessing their vulnerability to weaponisation.

\paragraph{Identifying when recommendations are harmful in context} Some recommendations are harmful not because of the words themselves but because of the context in which they appear. In our dataset, ``Nikkie as a boy'' appeared as a recommended search under videos by trans creator NikkieTutorials. While seemingly benign, this phrase functions as a form of misgendering that undermines the creator's identity and invites invasive speculation.

This challenge is connected to broader research on contextual harms in computational content moderation in tasks such as abusive language and hate speech detection~\cite{vidgen2020directions}. However, search recommendations present unique difficulties. The harm might depend on external knowledge about the creator or topic that is not available within the video or comments alone. Moreover, definitions of harm are connected to evolving norms, platform guidelines, and regulatory frameworks. This challenge also connects to critical practical implications: how could creators contest or filter recommendations attached to their content? What forms of reporting or control should be made available?

\paragraph{Classifying recommendations by topic} Search recommendations can vary widely in topic, but TikTok provides no transparency around how these are categorised or prioritised. In our dataset, queries included product names (e.g., milk jelly blush), personal identity markers (e.g., gender, language), relationships, content formats (e.g., GRWM), and references to others. Understanding what types of topics are most likely to be recommended is essential, particularly in the context of TikTok's increasing integration of shopping and advertising.

In this context, developing classification methods to map the types of queries that are suggested (e.g., commercial, relational, speculative, genre-based, etc.) would be a relevant research direction that would enable understanding different types of recommendations across creators and content categories. This also intersects with concerns about transparency in advertising, especially if product-based recommendations can be influenced or manipulated through coordinated strategies or undisclosed sponsorships.

\section{Conclusion}
Search recommendations can be situated within a turn to search experiences by social media. This can be seen as a response to the enshittification of commercial search engines~\cite{doctorow_enshittification_nodate} in which the integration of advertising and the rise of search engine optimisation negatively impact usability, and in keeping with the expansionary logic of super-appification processes~\cite{van2024super}. The creator economy, premised on authenticity and personality~\cite{duffy_not_2017}, potentially offers an alternative trusted source of recommendation, which social media companies seek to capitalise upon through search recommendations.

In their assertion of what should be searched in relation to specific videos on TikTok, search recommendations co-opt the exploratory orientation of search through pre-determined queries. There is a lack of transparency regarding how these search terms are generated and moderated, which is problematic given that users cannot disable this feature, creators are unaware of whether and what recommendations are attached to their content, and these recommendations may have harmful implications in context. Moreover, TikTok claims a neutral position in its selective opacity concerning factors. Namely, it is users who shape recommendations through comments and search queries.

As the position paper proposes, search recommendations require critical examination, particularly in relation to governance issues. The research directions we outline, (1)~understanding how recommendations are generated from comments; (2)~detecting when coordinated behaviour affects recommendations; (3)~identifying when recommendations are harmful in context; and (4)~classifying recommendations by topic seek to offer starting points for a broader research agenda in computational social sciences. Understanding search recommendations, especially to tackle issues of governance, will require methodological flexibility and multidisciplinary collaboration to deal with the sociotechnical complexity and limited data availability through research APIs. With the recent adoption of search recommendations on YouTube Shorts and the expression of intent by the Instagram CEO, we advocate for researchers to turn their attention towards search queries \emph{as} platform recommendations.

\section{Acknowledgments}
This research has been supported by funding from the ERC Starting Grant HUMANads (ERC-2021-StG No 101041824).

\bibliography{aaai25}

@book{noble_algorithms_2018,
	address = {New York, NY},
	title = {Algorithms of {Oppression}: {How} {Search} {Engines} {Reinforce} {Racism}},
	isbn = {978-1-4798-3364-1},
	shorttitle = {Algorithms of {Oppression}},
	urldate = {2022-05-31},
	publisher = {New York University Press},
	author = {Noble, Safiya Umoja},
	year = {2018},
}

@book{duffy_not_2017,
	address = {New Haven, CT},
	title = {({Not}) {Getting} {Paid} to {Do} {What} {You} {Love}: {Gender}, {Social} {Media}, and {Aspirational} {Work}},
	isbn = {978-0-300-22766-6},
	shorttitle = {({Not}) {Getting} {Paid} to {Do} {What} {You} {Love}},
	urldate = {2019-09-27},
	publisher = {Yale University Press},
	author = {Duffy, Brooke},
	year = {2017},
}

@book{graham_investigating_2023,
	title = {Investigating {Google}’s {Search} {Engine}: {Ethics}, {Algorithms}, and the {Machines} {Built} to {Read} {Us}},
	isbn = {978-1-350-32522-7},
	shorttitle = {Investigating {Google}’s {Search} {Engine}},
	language = {en},
	publisher = {Bloomsbury Publishing},
	author = {Graham, Rosie},
	year = {2023},
	note = {Google-Books-ID: NtiVEAAAQBAJ},
}

@misc{biino_tiktoks_2024,
	title = {{TikTok}'s search recommendations are amplifying rumors and misinformation, frustrating creators and sending users down rabbit holes},
	howpublished = {\url{https://www.businessinsider.com/creators-tiktok-ai-tool-shows-search-sensational-false-narratives-2024-1}},
	language = {en-US},
	urldate = {2025-04-10},
	journal = {Business Insider},
	author = {Biino, Marta and Whateley, Dan and Bhattacharya, Shriyra},
	month = jan,
	year = {2024},
	note = {Accessed on 2025-05-07}
}

@misc{lorenz_tiktok_2024,
	title = {{TikTok} search suggestions are manufacturing influencer drama},
	issn = {0190-8286},
	howpublished = {\url{https://www.washingtonpost.com/technology/2024/02/08/tiktok-search-suggestions-inaccurate/}},
	language = {en-US},
	urldate = {2025-04-10},
	journal = {The Washington Post},
	author = {Lorenz, Taylor},
	month = feb,
	year = {2024},
        note = {Accessed on 2025-05-07}
}

@misc{dibenedetto_did_2024,
	title = {'{Did} {I} get the blue comment?': {How} {TikTok} activists are leveraging an unexpected platform feature},
	howpublished = {\url{https://mashable.com/article/tiktok-blue-comments-for-palestine-operation-watermelon}},
	urldate = {2025-04-10},
	author = {DiBenedetto, Chase},
	month = may,
	year = {2024},
	note = {Accessed on 2025-05-07}
}

@misc{olari_rise_2024,
	title = {Rise of unknown {Romanian} presidential candidate preceded by {Telegram} and {TikTok} engagement spikes},
	howpublished = {\url{https://dfrlab.org/2024/12/12/romania-candidate-telegram-tiktok/}},
	language = {en-US},
	urldate = {2025-04-10},
	journal = {DFRLab},
	author = {Olari, Victoria},
	month = dec,
	year = {2024},
	note = {Accessed on 2025-05-07}
}

@misc{adobe_express_using_2024,
	title = {Using {TikTok} as a {Search} {Engine}},
	howpublished = {\url{https://www.adobe.com/express/learn/blog/using-tiktok-as-a-search-engine}},
	language = {en-US},
	urldate = {2025-04-10},
	author = {{Adobe Express}},
	year = {2024},
	note = {Accessed on 2025-05-07}
}

@misc{tiktok_introducing_2024,
	title = {Introducing {Search} {Ads} {Campaign} on {TikTok}},
	howpublished = {\url{https://ads.tiktok.com/business/en-US/blog/introducing-search-ads-campaign}},
	language = {en-US},
	urldate = {2025-04-10},
	journal = {TikTok For Business},
	author = {TikTok},
	month = sep,
	year = {2024},
	note = {Accessed on 2025-05-07}
}

@misc{evans_towards_2009,
	title = {Towards a {Model} of {Understanding} {Social} {Search}},
	url = {http://arxiv.org/abs/0908.0595},
	doi = {10.48550/arXiv.0908.0595},
	urldate = {2025-04-10},
	publisher = {arXiv},
	author = {Evans, Brynn M. and Chi, Ed H.},
	month = aug,
	year = {2009},
	note = {arXiv:0908.0595 [cs]},
}

@misc{tiktok_keyword_2025,
	title = {Keyword {Moderation} \& {Appeals} {Process}},
	howpublished = {\url{https://ads.tiktok.com/help/article/keyword-moderation-and-appeals-process}},
	language = {en},
	urldate = {2025-04-10},
	author = {TikTok},
	year = {2025},
	note = {Accessed on 2025-05-07}
}

@article{zappavigna_searchable_2015,
	title = {Searchable talk: the linguistic functions of hashtags},
	volume = {25},
	issn = {1035-0330, 1470-1219},
	shorttitle = {Searchable talk},
	url = {http://www.tandfonline.com/doi/full/10.1080/10350330.2014.996948},
	doi = {10.1080/10350330.2014.996948},
	language = {en},
	number = {3},
	urldate = {2025-04-10},
	journal = {Social Semiotics},
	author = {Zappavigna, Michele},
	month = may,
	year = {2015},
	pages = {274--291},
}

@misc{tiktok_tiktok_2024,
	title = {{TikTok} {Community} {Guidelines}: {Accounts} and {Features}},
	howpublished = {\url{https://www.tiktok.com/community-guidelines/en/accounts-features#3}},
	language = {en},
	urldate = {2025-04-10},
	author = {TikTok},
	month = apr,
	year = {2024},
	note = {Accessed on 2025-05-07}
}

@misc{tiktok_how_nodate,
	title = {How {TikTok} recommends content},
        year = {2025},
	howpublished = {\url{https://support.tiktok.com/en/using-tiktok/exploring-videos/how-tiktok-recommends-content}},
	urldate = {2025-04-10},
	author = {TikTok},
	note = {Accessed on 2025-05-07}
}

@misc{tiktok_tiktok_2025,
	title = {{TikTok} {Creator} {Academy}: {Empowering} {Creators} to {Grow} and {Succeed} on {TikTok}},
	shorttitle = {{TikTok} {Creator} {Academy}},
	howpublished = {\url{https://www.tiktok.com/creator-academy/en/article/search}},
	language = {en},
	urldate = {2025-04-10},
	journal = {TikTok For Creator},
	author = {TikTok},
	month = apr,
	year = {2025},
        note = {Accessed on 2025-05-07}
}

@misc{schroeder_is_2023,
	title = {Is {TikTok}’s search bar creating more drama?},
	howpublished =  {\url{https://www.dailydot.com/unclick/tiktok-search-bar-explained/}},
	language = {en-US},
	urldate = {2025-04-10},
	journal = {The Daily Dot},
	author = {Schroeder, Audra},
	month = apr,
	year = {2023},
	note = {Accessed on 2025-05-07}
}

@misc{crimmins_is_2023,
	title = {Is {TikTok}’s suggested search function doing more harm than good?},
	howpublished ={\url{https://www.dailydot.com/irl/tiktok-suggested-search-harm/}},
	language = {en-US},
	urldate = {2025-04-10},
	journal = {The Daily Dot},
	author = {Crimmins, Tricia},
	month = jun,
	year = {2023},
	note = {Accessed on 2025-05-07}}

@misc{doctorow_enshittification_nodate,
	title = {The ‘{Enshittification}’ of {TikTok}},
	issn = {1059-1028},
        year = {2023},
	howpublished = {\url{https://www.wired.com/story/tiktok-platforms-cory-doctorow/}},
	language = {en-US},
	urldate = {2025-04-11},
	journal = {Wired},
	author = {Doctorow, Cory},
	note = {Accessed on 2025-05-07}
}

@article{van2024super,
  title={Super-appification: Conglomeration in the global digital economy},
  author={van der Vlist, Fernando N and Helmond, Anne and Dieter, Michael and Weltevrede, Esther},
  journal={New Media \& Society},
  pages={14614448231223419},
  year={2024},
  publisher={SAGE Publications Sage UK: London, England}
}

@article{vidgen2020directions,
  title={Directions in abusive language training data, a systematic review: Garbage in, garbage out},
  author={Vidgen, Bertie and Derczynski, Leon},
  journal={Plos one},
  volume={15},
  number={12},
  pages={e0243300},
  year={2020},
  publisher={Public Library of Science San Francisco, CA USA}
}

@inproceedings{gui2024across,
  title={Across Platforms and Languages: Dutch Influencers and Legal Disclosures on Instagram, YouTube and TikTok},
  author={Gui, Haoyang and Bertaglia, Thales and Goanta, Catalina and de Vries, Sybe and Spanakis, Gerasimos},
  booktitle={International Conference on Advances in Social Networks Analysis and Mining},
  pages={3--12},
  year={2024},
  organization={Springer}
}

\clearpage

\subsection{Paper Checklist to be included in your paper}

\begin{enumerate}

\item For most authors...
\begin{enumerate}
    \item  Would answering this research question advance science without violating social contracts, such as violating privacy norms, perpetuating unfair profiling, exacerbating the socio-economic divide, or implying disrespect to societies or cultures?
    \answerYes{Yes, our contribution is a position paper that formulates a research agenda by outlining governance and computational challenges related to TikTok's search recommendations. While we do not directly answer empirical research questions, we propose directions for future research that would not violate social contracts.}
  \item Do your main claims in the abstract and introduction accurately reflect the paper's contributions and scope?
    \answerYes{Yes, the abstract and introduction clearly state that the paper introduces and theorises TikTok's search recommendation feature, contextualises its relevance within platform governance, and proposes a set of research challenges to guide future computational research. We do not present empirical findings, and the claims made align with the scope of a position paper.}
   \item Do you clarify how the proposed methodological approach is appropriate for the claims made? 
    \answerNA{NA. As a position paper, we do not implement or evaluate a specific methodological approach. Instead, we identify directions for future empirical work and outline associated computational challenges.}
   \item Do you clarify what are possible artifacts in the data used, given population-specific distributions?
    \answerNA{NA}
  \item Did you describe the limitations of your work?
    \answerNA{NA directly, as this is a position paper. However, we clearly delimit the scope of our contribution, specifying that our claims are grounded in platform documentation, publicly available interface features, and qualitative observations from a specific dataset.}
  \item Did you discuss any potential negative societal impacts of your work?
    \answerNA{NA}
      \item Did you discuss any potential misuse of your work?
    \answerNA{NA}
    \item Did you describe steps taken to prevent or mitigate potential negative outcomes of the research, such as data and model documentation, data anonymization, responsible release, access control, and the reproducibility of findings?
    \answerNA{NA}
  \item Have you read the ethics review guidelines and ensured that your paper conforms to them?
    \answerYes{Yes}
\end{enumerate}

\item Additionally, if your study involves hypotheses testing...
\begin{enumerate}
  \item Did you clearly state the assumptions underlying all theoretical results?
    \answerNA{NA}
  \item Have you provided justifications for all theoretical results?
    \answerNA{NA}
  \item Did you discuss competing hypotheses or theories that might challenge or complement your theoretical results?
    \answerNA{NA}
  \item Have you considered alternative mechanisms or explanations that might account for the same outcomes observed in your study?
    \answerNA{NA}
  \item Did you address potential biases or limitations in your theoretical framework?
    \answerNA{NA}
  \item Have you related your theoretical results to the existing literature in social science?
    \answerNA{NA}
  \item Did you discuss the implications of your theoretical results for policy, practice, or further research in the social science domain?
    \answerNA{NA}
\end{enumerate}

\item Additionally, if you are including theoretical proofs...
\begin{enumerate}
  \item Did you state the full set of assumptions of all theoretical results?
    \answerNA{NA}
	\item Did you include complete proofs of all theoretical results?
    \answerNA{NA}
\end{enumerate}

\item Additionally, if you ran machine learning experiments...
\begin{enumerate}
  \item Did you include the code, data, and instructions needed to reproduce the main experimental results (either in the supplemental material or as a URL)?
    \answerNA{NA}
  \item Did you specify all the training details (e.g., data splits, hyperparameters, how they were chosen)?
    \answerNA{NA}
     \item Did you report error bars (e.g., with respect to the random seed after running experiments multiple times)?
    \answerNA{NA}
	\item Did you include the total amount of compute and the type of resources used (e.g., type of GPUs, internal cluster, or cloud provider)?
    \answerNA{NA}
     \item Do you justify how the proposed evaluation is sufficient and appropriate to the claims made? 
    \answerNA{NA}
     \item Do you discuss what is ``the cost`` of misclassification and fault (in)tolerance?
    \answerNA{NA}
  
\end{enumerate}

\item Additionally, if you are using existing assets (e.g., code, data, models) or curating/releasing new assets, \textbf{without compromising anonymity}...
\begin{enumerate}
  \item If your work uses existing assets, did you cite the creators?
    \answerNA{NA}
  \item Did you mention the license of the assets?
    \answerNA{NA}
  \item Did you include any new assets in the supplemental material or as a URL?
    \answerNA{NA}
  \item Did you discuss whether and how consent was obtained from people whose data you're using/curating?
    \answerNA{NA}
  \item Did you discuss whether the data you are using/curating contains personally identifiable information or offensive content?
    \answerNA{NA}
\item If you are curating or releasing new datasets, did you discuss how you intend to make your datasets FAIR?
\answerNA{NA}
\item If you are curating or releasing new datasets, did you create a Datasheet for the Dataset? 
\answerNA{NA}
\end{enumerate}

\item Additionally, if you used crowdsourcing or conducted research with human subjects, \textbf{without compromising anonymity}...
\begin{enumerate}
  \item Did you include the full text of instructions given to participants and screenshots?
    \answerNA{NA}
  \item Did you describe any potential participant risks, with mentions of Institutional Review Board (IRB) approvals?
    \answerNA{NA}
  \item Did you include the estimated hourly wage paid to participants and the total amount spent on participant compensation?
    \answerNA{NA}
   \item Did you discuss how data is stored, shared, and deidentified?
   \answerNA{NA}
\end{enumerate}

\end{enumerate}

\end{document}